**Symmetry-selective quasiparticle scattering and electric field tunability of the ZrSiS surface electronic structure**

Michael S. Lodge[1,2,3,‡], Elizabeth Marcellina[1,†,‡], Ziming Zhu[4], Xiao-Ping Li[5], Dariusz Kaczorowski[6], Michael S. Fuhrer[7,8], Shengyuan A. Yang[5], Bent Weber[1,8,*]

1   School of Physical and Mathematical Sciences, Nanyang Technological University, 637371, Singapore
2   Department of Physics, University of Central Florida, Orlando, Florida 32816, United States
3   NanoScience Technology Center, University of Central Florida, Orlando, Florida 32826, United States
4   Key Laboratory of Low-Dimensional Quantum Structures and Quantum Control of Ministry of Education, Department of Physics and Synergetic Innovation Center for Quantum Effects and Applications, Hunan Normal University, Changsha 410081, China
5   Research Laboratory for Quantum Materials, Singapore University of Technology and Design, 487372 Singapore
6   Institute of Low Temperature and Structure Research, Polish Academy of Sciences, Okólna 2, 50-422 Wrocław, Poland
7   School of Physics and Astronomy, Monash University, Clayton VIC 3800 Australia Monash Centre for Atomically Thin Materials, Monash University, Clayton VIC 3800 Australia
8   Australian Research Council (ARC) Centre of Excellence for Low-Energy Electronics Technologies (FLEET), School of Physics, Monash University, Clayton VIC 3800 Australia

‡These authors contributed equally.

**Corresponding Authors**

*e-mail: b.weber@ntu.edu.sg (main corresponding author)

†e-mail: elizabeth.marcellina@sydney.edu.au (co-corresponding author). Present address: School of Physics, The University of Sydney, New South Wales 2006, Australia.

**ABSTRACT**

3D Dirac semimetals with square-net non-symmorphic symmetry, such as ternary ZrXY (X=Si, Ge; Y=S, Se, Te) compounds, have attracted significant attention owing to the presence of topological nodal lines, loops, or networks in their bulk. Orbital symmetry plays a profound role such materials as the different branches of the nodal dispersion can be distinguished by their distinct orbital symmetry eigenvalues. The presence of different eigenvalues suggests that scattering between states of different orbital symmetry may be strongly suppressed. Indeed, in ZrSiS, there has been no clear experimental evidence of quasiparticle scattering between states of different symmetry eigenvalue has been reported at small wave vector $\vec{q}$. Here we show, using quasiparticle interference (QPI), that atomic step-edges in the ZrSiS surface facilitate quasiparticle scattering between states of different symmetry eigenvalues. This symmetry eigenvalue mixing quasiparticle scattering is the first to be reported for ZrSiS and contrasts quasiparticle scattering with no mixing of symmetry eigenvalues, where the latter occurs with scatterers preserving the glide mirror symmetry of the crystal lattice, e.g., native point defects in ZrSiS. Finally, we show that the electronic structure of the ZrSiS surface, including its unique floating band surface state (FBSS), can be tuned by a vertical electric field locally applied by the tip of a scanning tunneling microscope (STM), enabling control of a spin-orbit induced avoided crossing near the Fermi level by as much as 300%.

**INTRODUCTION**

Three-dimensional (3D) topological semimetals feature symmetry-protected band crossings [1-5], classified into Dirac/Weyl nodes, nodal lines, rings, or knots [6], with linear band dispersion at low energy. The presence of relativistic (Dirac/Weyl) fermions has been demonstrated to give rise to extremely high electron mobility [7-9], non-saturating magnetoresistance [9-14], chiral anomaly [15-19], anomalous Hall effect [20], large photocurrents [21-26], anomalous Nernst effect [27-29], planar Hall effect [30-34], and negative longitudinal magnetoresistance [15, 35-41]. The wide range of band topology- and spin-related phenomena found in topological semimetals may open new avenues towards low-dissipation electronic and spintronic devices, and efficient photodetectors [42]. Of particular interest is the family of square-net ternary compounds ZrXY (X = Si [43, 46-60], Ge [61, 62]; Y = S, Se, Te) and HfSiS [43-45]. In these materials, the non-symmorphic symmetry gives rise to Dirac nodal lines, points and loops, which have been considered for their potential roles in spintronics and spin caloritronics [63, 64].

ZrXY compounds are also known to exhibit strong spin-orbit coupling (SOC). For example, ZrSiS has been predicted to possess a large spin Hall conductivity of -611 ℏ/e, which is only an order of magnitude lower compared to that of Pt (of 2200 ℏ/e) [63]. Such large spin Hall conductivities could generate spin currents without the need for an external magnetic field [63]. More recently, Shubnikov-de Haas oscillation measurements have revealed that the spin-orbit interaction in ZrSiS gives rise to an anisotropic Berry phase as large as $0.15\pi$ [65]. Furthermore, in heavier square net materials (such as in ZrGeTe), the even stronger SOC gives rise to additional band inversions, hence forming additional Dirac nodes [62]. The strong SOC in ZrXY-type materials, together with the nodal line dispersion, thus provides a fertile ground for investigations on the interplay between spin-orbit coupling, symmetry, and topological phases.

In this work, we perform quasiparticle interference (QPI) spectroscopy resolved by scanning tunneling microscopy (STM) at 4.5 K to probe the surface states of ZrSiS and demonstrate a new QPI scattering route in ZrSiS that arises from symmetry eigenvalue mixing. In QPI, extended electronic states are scattered off crystal defects, thereby mixing states with momenta $\vec{k}_i$ and $\vec{k}_f$ and giving rise to a modulation in the local density of states (LDOS) with frequency $2\pi/|\vec{q}|$, where $|\vec{q}|=|\vec{k}_f - \vec{k}_i|$ [66, 67]. Owing to the sensitivity of QPI interference patterns to the scattering wave vectors, as well as their high energy resolution, Fourier transform scanning tunneling spectroscopy (FT-STM) [68] serves as a powerful tool to resolve a material's energy dispersion (e.g., 3D Dirac nodal line dispersions of ZrSiS and ZrSiSe [52-57]) and provides direct evidence for the symmetry- or topological protection of electronic states [69-71]. After using QPI to show symmetry-selective scattering of the ZrSiS states, we additionally show that the electronic structure of the ZrSiS surface, including its "floating band" surface state (FBSS) and a spin-orbit coupling induced band anticrossing, can be tuned by a vertical electric field applied by the STM tip. Specifically, we exploited the tip-induced field effect to widen the spin-orbit gap by nearly 300%.

**MATERIALS AND METHODS**

Bulk crystals (~ 1 mm$^3$) of ZrSiS grown by chemical vapor transport method [12, 72] were cleaved in UHV using a ceramic cleaving post, immediately prior to loading into the STM cryostat. The experiments were then performed at liquid helium temperature ($T \cong 4.5$ K) in both a Createc and an Omicron low-temperature STM system. The STM's Pt/Ir tip was prepared by repeated controlled indentations into a herringbone-reconstructed Au(111) surface, which provides the tip's apex with Au coating [73]. All tips were calibrated in STS against the known spectral features of the Au(111) Shockley surface state [74]. The differential conductance for QPI and tip height dependence data were measured using standard lock-in technique at $f = 747$ Hz, $V_{RMS} = 20$ mV and $V_{RMS} = 5$ mV and $f = 707$ Hz, respectively. Large area QPI from defect ensembles was measured in constant current mode to mitigate low-frequency noise, while QPI at the step edge was measured in constant height mode.

**RESULTS AND DISCUSSION**

*Quasiparticle scattering by atomic site defects*

A detailed ZrSiS crystal and electronic structure is summarized in Fig. 1. Fig. 1A shows the Brillouin zones (BZ) for the 3D bulk and its 2D surface projection ($k_z = 0$) (white dashed lines), highlighting the position of the nodal lines in momentum space, the high symmetry directions, and lattice constants *a*, *b*, and *c*. Closed nodal loops at bulk and surface ensue from the crossings of bands with different orbital character, and hence, different symmetry eigenvalues [45, 47, 48]. The relationship between symmetry eigenvalues to the Dirac bands and possible scattering wave vectors are illustrated in Figs. 1B, C. Fig. 1B shows simplified schematics of the nodal dispersion along $\bar{\Gamma} - \bar{M}$, while Fig. 1C shows the corresponding constant energy contours at an energy above a Dirac nodal line. According to DFT calculations [48], the "inner" branch/square of the nodal dispersion (closer to $\bar{\Gamma}$) is composed of the Zr $d_{xz}/d_{yz}$ orbital, while the "outer" branch/square stems from the Zr $d_{x^2-y^2}$ orbital. Furthermore, the distinct orbital character of these bands can be described by distinct symmetry eigenvalues ($\nu = 1$ for the $d_{xz}/d_{yz}$ orbital and $\nu = 0$ for the $d_{x^2-y^2}$ orbital), which impose stringent selection rules for quasiparticle scattering [48].

The electronic band dispersion as obtained from our density functional theory (DFT) calculation of a 3-layer slab is shown in Fig. 1D, clearly displaying Dirac nodes close to the Fermi level ($E = 0$) in the band manifolds along the $\bar{\Gamma} - \bar{M}$ and $\bar{\Gamma} - \bar{X}$ directions. Green-coloured bands carry a large spectral weight at the ZrSiS surface, which the STM is predominantly sensitive to. While the bulk band structure including orbital symmetry are reflected at the surface, breaking of the non-symmorphic symmetry across the vertical glide-mirror planes shifts the bulk bands (BB) in energy, including the $k_z = 0$ nodal loop. Symmetry breaking at the surface further lifts the fourfold band degeneracy of the bulk Dirac node near the $\bar{X}$ point [48, 50, 51, 53-55, 57], shifting the surface bands downward in energy by approximately ~1 eV. This gives rise to a saddle-like state with a large spectral weight in the surface (green lines in Fig. 1D) which extends across $\bar{\Gamma} - \bar{X} - \bar{M}$. As this saddle-like state is completely unpinned from corresponding bulk bands at certain energies and wave vectors, it has been termed a "floating band" surface state (FBSS) [47, 55, 56], previously identifiable in ARPES [47, 50], QPI [52-57] and transport [9, 55, 59, 60] experiments.

The FBSS "saddle" with concomitant band flattening near $\bar{X}$ is reflected (Fig. 1E) in an enhancement of the local density of states (LDOS) near -300 meV (red arrow), measured in STM/STS at T = 4.5 K, and in good agreement with the projected density of states (PDOS) calculated. The expected double-peak structure in the LDOS, owing to the two inflection points near the FBSS saddle, is more clearly observed with increased spectral resolution in Fig. 3B (discussion below). The minimum at $E$ = -21 meV reflects the reduction in the measured LDOS (black arrow) for energies near the Dirac nodal line (DNL). Energies $E$ > 136 meV corresponds to the onset of parabolic bulk bands, reflected as a strong enhancement of the LDOS.

We then probe the surface states of ZrSiS, including the FBSS using QPI spectroscopy [52-57, 66-71, 75-81] (Figs. 1G-I). A comparison of an atomic-resolution STM image of the S-terminated ZrSiS surface after cleaving and its corresponding differential conductance (d$I$/d$V$) map at $(E - E_F)$ = +500mV is shown in Figs. 1F, G. The d$I$/d$V$ map has been processed by Fourier filtering of the atomic lattice to highlight QPI induced charge density modulations with wave vector $\vec{q} = \vec{k}_f - \vec{k}_i$. The dominant scattering vectors $\vec{q}$ and hence the likelihood of scattering between different initial and final states can be obtained from a 2D Fast Fourier Transform (FFT), which are then compared with the corresponding surface spectral weight at a similar energy in Figs. 1H, I. We find that scattering within the surface floating band [54, 56, 57] gives rise to a square pattern at small wave vector ($\vec{q}_1$) (Fig. 1I). Transitions between states within the concentric squares in the $\bar{\Gamma} - \bar{M}$ direction are most prominently observed in $\vec{q}_4, \vec{q}_5$ and, to a lesser extent, $\vec{q}_3$. The vector $\vec{q}_6$ corresponds to scattering between the surface and bulk [57] bands in the $\bar{\Gamma} - \bar{X}$ direction. Notably, scattering with small momentum transfer between the concentric squares (Fig. 1B), and hence between states of different symmetry eigenvalues, is strongly suppressed as inferred from the absence of spectral feature corresponding to scattering vector $\vec{q}_2$ (dashed red square in Fig. 1I [54]). This indicates that native atomic point defect [56] and defect ensembles do not allow for scattering between states of different orbital symmetry eigenvalues [48, 57]. This finding is consistent with the lack of any report of a clear, unambiguous signature of this scattering vector across all QPI studies published so far [53-57] and confirms that orbital symmetry is effective in forbidding these states from quasiparticle scattering.

*Quasiparticle scattering by atomic line defect*

After investigating QPI with point defects as scatterers, we then perform QPI measurements at an atomic step edge of our ZrSiS sample. Step edges have different a spatial symmetry from point defects given that the former break translational invariance with a strong short range scattering potential and impose a hardwall boundary on surface-confined state. Indeed, atomic step edges have been extensively investigated for the presence of 1D electronic edge states [77-84] as well as scattering/transmission of 2D surface states, e.g. on the surfaces of 3D topological insulators (TIs) [81, 84], Weyl [87] and Dirac semimetals [88], as well as higher-order topological insulators [77, 89]. For 3D TI surface states, it has been shown that translational symmetry breaking within the surface [86] faciliates scattering of the otherwise topologically protected 2D helical state.

In Fig. 2A, we show the topography around a ZrSiS surface line defect atomic step, terminated along the [110] direction of the crystal, allowing us to probe the $\bar{\Gamma} - \bar{M}$ direction in reciprocal space (see inset). An STM height profile and a corresponding series of point spectra along a line (arrow in Fig. 2A) across the edge are shown in Figs. 2B and 2C. To highlight the charge density modulations due to quasi-particle scattering and interference, we have subtracted a smooth background reference spectrum, taken from the upper terrace at a distance 10 nm from the edge, where the modulations subside (Supplementary Fig. 1). We find a rich QPI pattern on the upper ZrSiS terrace (Fig. 2C), which is nearly absent on the lower terrace. To identify the dominant scattering vectors, we compare 1D fast Fourier transforms (FFTs) for both the lower and the upper terraces (Figs. 2E, F, and Fig. S2 of the Supplementary Material), with the corresponding DFT dispersion in the $\bar{M} - \bar{\Gamma} - \bar{M}$ direction (Fig. 2D). Expected scattering vectors are indicated by white arrows, and are in excellent agreement with those determined in Figs. 1H, I. The absence of any prominent features in the lower terrace FFT confirms that scattering by the step edge is strongly suppressed here, with only indications of one scattering vector ($\vec{q}_7$) resolved within the noise, which reduces to $|\vec{q}| = 0$ at $E \approx$ -200 meV, reflecting the DNL in the ZrSiS bulk. The complex QPI pattern on the upper terrace allows us to identify as many as four scattering vectors ($\vec{q}_2$, $\vec{q}_4$, $\vec{q}_5$ and $\vec{q}_7$), with $\vec{q}_{2,4,5}$ being in excellent agreement with those predicted in Fig. 1H, and $\vec{q}_{4,5}$ with previous reports [52-57].

However, in stark contrast to the absence of scattering vector $\vec{q}_2$ on terraces without symmetry eigenvalue mixing (Fig. 1H,I, Fig. 2E, as well as previous reports [52-57]), we find that $\vec{q}_2$ is a very noticeable scattering vector at the upper terrace near the step edge (Fig. 2F), with comparable prominence to $\vec{q}_4$ and $\vec{q}_5$ (see also Fig. S2 of the Supplementary Material). The vector $\vec{q}_2$ corresponds to small momentum scattering between states of different symmetry eigenvalues, i.e., the step edge effectively mixes the Zr $d_{xz}/d_{yz}$ ($v = 1$) and $d_x{}^2\text{-}{}_y{}^2$ ($v = 0$) orbitals. To our knowledge, this is the first time [48, 57] that this scattering vector $\vec{q}_2$ has been unambiguously identified in ZrSiS. Previously, the were faint signatures of $\vec{q}_2$ in ZrSiSe [57], but the observation has not been discussed in the context of symmetry eigenvalues. Additionally, to the best our knowledge, there have not been other reports of QPI scattering in ZrSiX-type compounds in which $\vec{q}_2$ is identified.

The scattering vector $\vec{q}_2$ further allows us to confirm the presence of the Dirac nodal dispersion at the surface and extract the nodal line energy of $E = 140$ meV in the $\bar{\Gamma} - \bar{M}$ direction, where $|\vec{q}_2|$ reduces to zero. We note that, although vectors $\vec{q}_2$ and $\vec{q}_4$ represent scattering between states of different symmetry eigenvalues, there are important differences. The vector $\vec{q}_2$ corresponds to a small wave vector scattering on the same side of the Brillouin zone, whereas $\vec{q}_4$ corresponds to a large wave vector scattering across the opposite sides of the Brillouin zone [54-57]. The non-dispersive features in Fig. 2E may have originated from symmetry eigenvalue mixing that could occur at energies far above or below the nodal line even in clean terraces [52, 54, 57]. Furthermore, since the vectors $\vec{q}_4$ (which mixes different symmetry eigenvalues) and $\vec{q}_5$ (which does not mix different symmetry eigenvalues) are closely separated in wave vector (by ~$0.1 \times \frac{2\pi}{a}$) and can be difficult to distinguish (compare with Fig. 1I), $\vec{q}_2$ is the only unambiguous proof of a QPI scattering that mixes different symmetry eigenvalues.

*Electric field tuning of the floating bands and spin-orbit gap*

Following identification of the ZrSiS surface nodal dispersion, we proceed to showing that the ZrSiS surface electronic structure can be effectively tuned by applying a strong vertical electric field locally via the STM tip [90]. Local field-effect tuning of semiconductor surfaces, often referred to as tip-induced bend bending, is a well-established phenomenon on metal-insulator-semiconductor junctions [91-93], but more recently has also been used to field-tune the band structure of atomically thin layers of the 3D Dirac semimetal $Na_3Bi$ [90]. In Ref. [90], tip-induced electric fields have been shown to Stark shift electronic bands near the Fermi energy to induce a topological phase transition [90]. As shown schematically in Fig. 3A, tip-induced band bending occurs when the STM tip is placed in proximity to a sample of a different work function. The difference in work function $\Delta\phi$ combined with the applied bias then causes a built-in electric field $F_z = -\frac{\partial U}{\partial z} = \Delta\phi/z$ at the junction, where $U$ and $z$ are the electric potential and the tip height, respectively.

Fig. 3B shows the measured tunnelling LDOS as a function of the tip-sample separation $\Delta z = z - z_0$, measured at position $z$ relative to the initial tip height $z_0$, similar to the approach of Ref. [90]. For each spectral feature identified in Fig. 3B, there is a monotonic shift in energy as the tip approaches the sample surface (i.e., $\Delta z$ becomes more negative). Of particular interest are two LDOS peaks found at -158 meV and -323 meV (arrows) when $\Delta z = 0$, which have previously been attributed to arise from the flattened bands at the inflection points of the FBSS saddle [50-53, 55]. These FBSS shift lower in energy by as much as 200 meV as the tip approaches the surface, which we interpret as evidence for electric field tunability of the floating band state. This interpretation is indeed supported by our DFT calculations (Fig. 3B, righthand side, blue lines) for zero and finite ($F_z = 1$ V Å$^{-1}$) vertical electric field (red lines), respectively, which reproduce the observed surface band shifts (black arrows in Fig. 3B) well. Our calculations further confirm that at a strong vertical electric field breaks the degeneracy of the FBSS between top (bottom) surfaces, whereby the surface bands shift down (up) in energy, respectively. The polarity of the electric field is also confirmed through the dependence of surface-concentrated parabolic bulk bands (BB) between $\bar{\Gamma} - \bar{X}$ on the applied electric field, where the bands shift upward in energy in both experiment and DFT calculations.

We obtain a rough estimate of the built-in electric field strength by considering the work function difference at the junction, following the procedure in Ref. [90]. As the work function of ZrSiS is unknown, we first extract an average work function of tip (*t*) and sample (*s*) $\bar{\phi} = \frac{1}{2}(\phi_s+\phi_t)$ by fitting $I(z) = I_0 e^{-2\kappa z}$ to the exponential dependence of the tunneling current in Fig. 3B, where $\kappa = \sqrt{2m\bar{\phi}}/\hbar$ and *m* the free electron mass. We find $\kappa = 1.04$ Å$^{-1}$ and $\bar{\phi} = 4.16$ eV. Further assuming the work function of a gold-coated Pt/Ir tip, ($\phi_{tip}$ = 5.3 eV [92]), we estimate $\phi_s$ =3.02 eV for ZrSiS, comparable with the work functions of both Zr and Si [94]. Hence, for a typical tip-sample separation of $z$ = 1 nm and work function difference $\Delta\phi = \phi_s - \phi_t$ = 2.28 eV, we estimate an electric field of $F_z \approx 2$ V/nm applied at zero bias. We do note that this electric field estimate is substantially smaller compared to the assumptions made in the theoretical calculations. However, this discrepancy is consistent with previous demonstrations of the tip-induced electric field effect [90]. Also, our simplified estimate of the built-in electric field relies on the assumption of a planar tip-sample junction and should thus be considered a *lower* bound to the field strength. Indeed, the atomically sharp probe of the STM may be expected to lead to a considerable enhancement of the electric field strength in the region directly underneath the tip's apex where the LDOS probed [93].

Finally, we focus on the narrow suppression of the LDOS observed in Fig. 3B near $E = E_F$. A close-up of this feature and its tip-height dependence is shown in Fig. 3C. The position of the LDOS minimum $V_c$ of the suppression and its width $\Delta$ are plotted in Fig. 3D, respectively, as extracted from Gaussian fits (solid lines in Fig. 3C). We observe a clear shift of $V_c$ and widening of the LDOS suppression as the tip approaches the surface (solid and dashed lines are guides to the eye). Comparing to our DFT model (Fig. 3B), we interpret the feature observed in Fig. 3C as a spin-orbit induced avoided crossing that causes the formation of a soft energy gap to near the $\bar{X}$-point (see inset to Fig. 3B comparing DFT with and without SOC). We find that $\Delta$ can be tuned between 30 meV and 85 meV, nearly tripling over the range of tip heights probed. For the asymptotic limit at a large tip-sample distance (and hence zero electric fields), we extract a spin-orbit gap $\Delta = 20$ meV, in good agreement with our DFT predictions ($\Delta = 11.6$ meV), as well as with previous ARPES-based measurements ($\Delta = 10$ meV) [50, 51], confirming that the change in the spin-orbit gap in our sample is indeed due to the applied electric field.

## SUMMARY

To summarize, we have resolved quasiparticle scattering and interference within the surface of the non-symmorphic nodal-line Dirac semimetal ZrSiS. Our results confirm the theoretical predictions on the symmetry eigenvalue-dependent selection rules for QPI scattering in non-symmorphic layered materials [31]. While the QPI scattering between nodal dispersion states of different symmetry eigenvalues is forbidden for weak scatters given by native atomic point defect ensembles in the surface, an atomic step edge provides an avenue for quasiparticle scattering between states of different symmetry eigenvalue at small wave vector $\vec{q}_2$. Such symmetry eigenvalue mixing is the first time to be reported for ZrSiS. Furthermore, the small wave vector QPI scattering further allows us to resolve the ZrSiS nodal line dispersion. Finally, we have shown that the ZrSiS surface band structure, including its signature floating band surface states, can be tuned by a strong vertical electric field applied locally by an STM tip. In agreement with DFT calculations, we find that a spin-orbit (SOC) induced band anticrossing ($\Delta = 20.35$ meV) at the $\bar{X}$-point of the Brillouin zone, can be enhanced threefold by the local field effect. We believe that our work may stimulate further investigations on the interplay between symmetry, spin-orbit coupling, and topological phases in non-symmorphic square-net semimetals.

# FIGURES

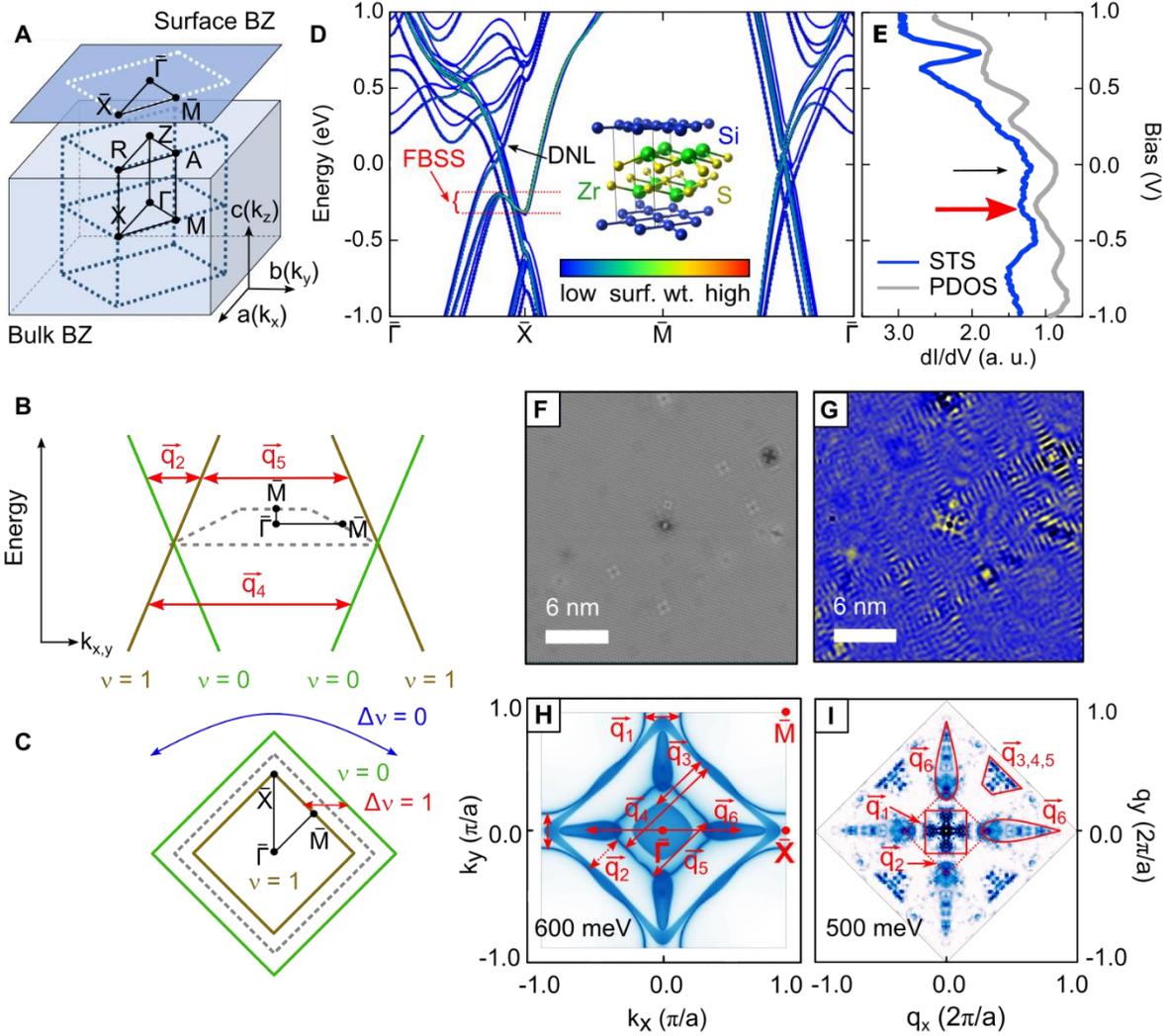

**Figure 1: Atomic and electronic structure of the nodal loop semimetal ZrSiS.** (**A**) Brillouin zones of the ZrSiS surface and bulk. (**B, C**) Schematic of the Dirac nodal line dispersion distinguishing the different symmetry eigenvalues (green and brown solid lines) and indicating the position of the nodal line (black dashed lines). (**B**) illustrates the scattering vectors along the $\bar{\Gamma} - \bar{M}$ direction between the Dirac bands. For simplicity, the spin-orbit interaction induced gap at $E \approx 0$ eV near the $\bar{X}$ point is not shown in (**B**). (**C**) is a simplified constant energy contour of (**B**), where scattering between the same ($\Delta \nu = 0$) and different symmetry eigenvalues ($\Delta \nu = 1$) is illustrated. The dashed line corresponds to the Fermi contour at the nodal line energy whilst the solid lines are the Fermi contours above the nodal line energy. (**D**) Calculated surface (slab, 3UC) band structure of ZrSiS along the high symmetry directions, without spin-orbit coupling. **Inset:** Crystal structure of ZrSiS. (**E**) Measured d$I$/d$V$ spectrum of pristine ZrSiS (blue) compared against the calculated PDOS (gray) for an S-terminated surface, showing the energies of the bulk bands (BB), Dirac nodal line (DNL), and saddle-like state (SS) bands. (**F**) STM topography image of as-cleaved ZrSiS showing a S-terminated surface with atomic-site defects, acquired with sample bias $V = 500$ mV and current setpoint $I_T = 300$ pA at $T = 4.8$ K. (**G**) d$I$/d$V$ map of the sample region in (**F**). (**H**) Slab calculation (3 UC) of surface constant energy contour at $E = 600$ meV showing scattering between floating surface states. (**I**) Symmetrized 2D FFT of the quasiparticle interference from a larger d$I$/d$V$ map from which (**G**) as taken.

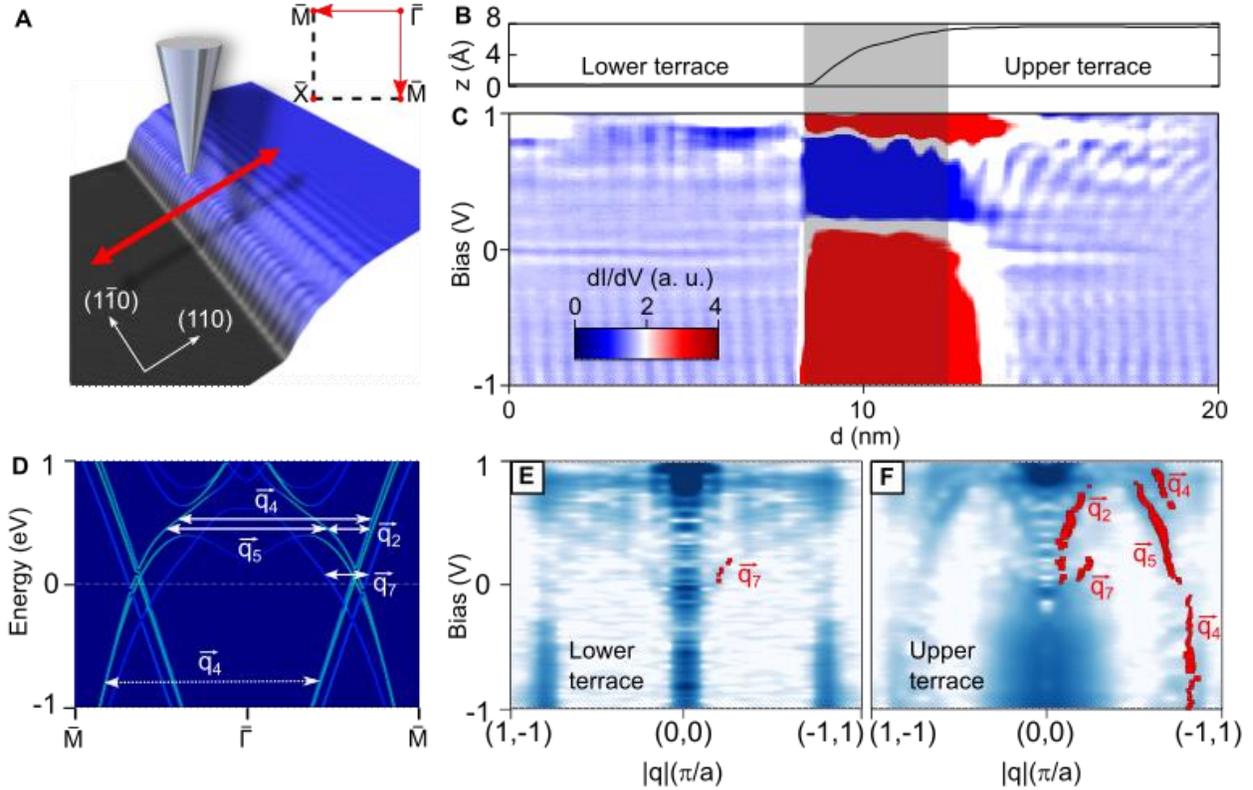

**Figure 2: Quasiparticle scattering and interference from an atomic step edge.** (**A**) A 3D topography image of a ZrSiS step edge, running across the [110]-crystal direction, superimposed with its d$I$/d$V$ map to highlight the charge density modulations observed. The step edge provides a strong short-range scattering potential giving rise to quasiparticles scattering and interference. (**Inset**) Equivalent $\bar{\Gamma} - \bar{M}$ direction in momentum space (red arrow) in which the QPI scattering vectors are resolved. (**B**) Topographic height-profile across the step edge, showing the upper and lower terraces. (**C**) Spatially resolved (with a 0.67 Å spacing) d$I$/d$V$ spectra along a 20 nm line in the [110] ($\bar{\Gamma} - \bar{M}$) direction across the step edge (along the red arrow in A), with reference background subtracted. The lock-in parameters used for the spectra were $f$ = 747 Hz and $V_{RMS}$ = 20 mV. (**D**) Calculated surface band structure along the $\bar{\Gamma} - \bar{M}$ direction with the corresponding quasiparticle scattering vectors $\vec{q}$ indicated. (**E, F**) 1D row-wise FFT of the QPI data in (**C**), showing the QPI dispersion for (**E**) the lower terrace and (**F**) the upper terrace, with energy-$\vec{q}$ dispersions from (**D**) indicated (see also Fig. S2 of the Supplementary Material). The Fourier transform excludes a narrow region across the step edge (gray shaded box in Figs. 2**B, C**) to remove possible artifacts arising from the tip's finite radius as the tip crosses the step edge. The red markers' positions indicate the FFT peak positions for $\vec{q}_2$, $\vec{q}_4$, $\vec{q}_5$ and $\vec{q}_7$, whereas the size of the red markers is directly proportional to the amplitude of the FFT peaks. The topographic data in (**A**) was acquired with sample bias $V$ = 600 mV and tunneling current setpoint $I$ = 300 pA.

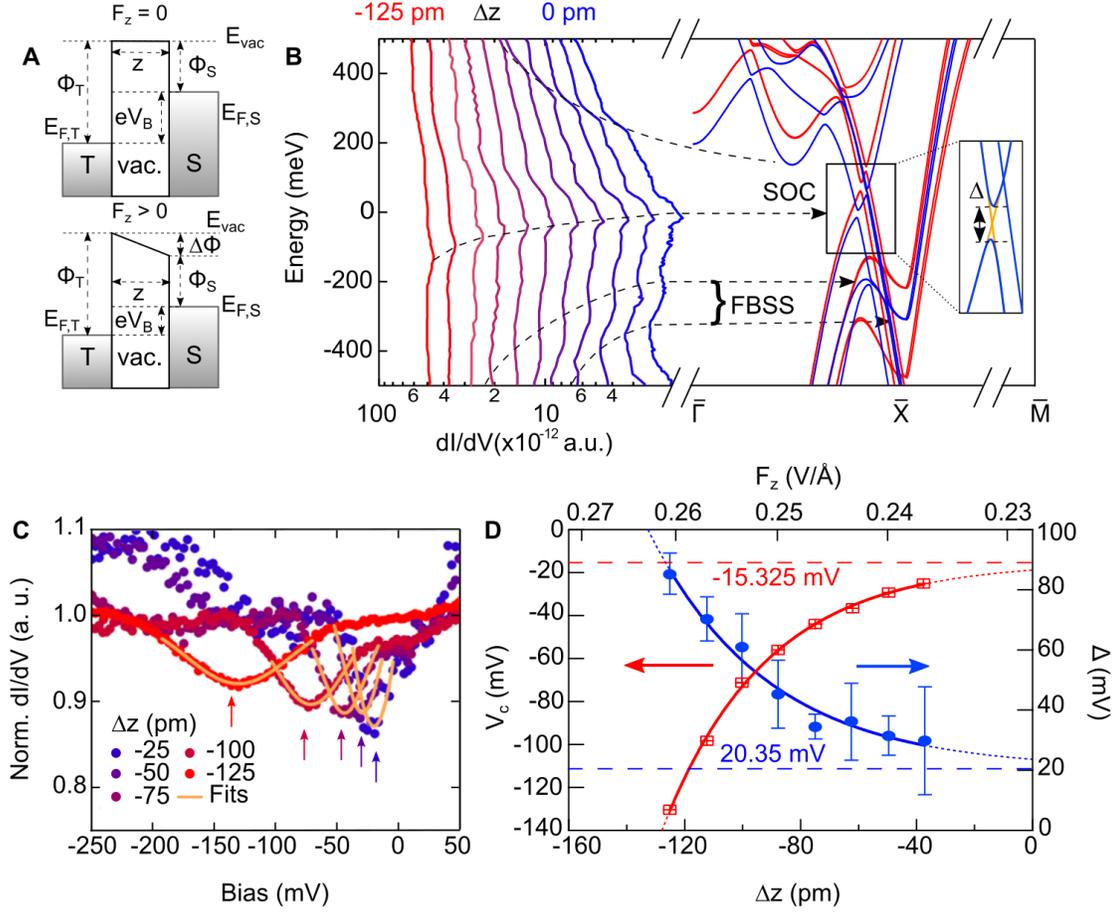

**Figure 3: Electric field tunability of the ZrSiS surface electronic structure.** (**A**) Schematic of the energy diagram for a tip-vacuum-metallic sample system, illustrating how electric fields $F_z = 0$ V/Å (top) and $F_z > 0$ V/Å (bottom) vary across the junction. Here, $V_B$ is the bias, $F_z$ is the electric field between tip and sample, $\Phi_{T(S)}$ is the tip (sample) work function, $\Delta\Phi$ is the difference between the tip and sample work functions, $E_{F,T(S)}$ are the tip (sample) Fermi energy, and $E_{vac}$ is the vacuum energy level. (**B**) (Left) d$I$/d$V$ point spectra as a function of changes in the tip-sample distance for increments of $|\Delta z| = -12.5$ pm. The spectra were acquired using lock-in parameters $V_{RMS} = 5$ mV and $f = 707$ Hz. (Right) DFT slab calculation (3 UC) of the surface band structure of ZrSiS with (red) and without (blue) an electric field of 1 eV/Å applied. In both cases, SOC was included in the calculations. Insert: Close-up of the SOC-induced energy gap $E_g$ (=11.6 meV) due to spin-orbit coupling (blue), that is absent without spin-orbit coupling (orange). The dashed lines are guides to the eye. (**C**) Close-up of the measured SOC-induced gap in d$I$/d$V$ (normalized). Solid lines are Gaussian fits. (**D**) Parameters extracted from the Gaussian fits to (**C**), indicating the gap center ($V_c$) and the width ($\Delta$). In the asymptotic limit of the tip being far from the sample (exponential fits), we find $V_c = -15.325$ meV and $\Delta = 20.35$ mV, respectively.

**AUTHOR INFORMATION**

**Author Contributions**

MSL and BW performed the scanning tunnelling microscopy and spectroscopy experiments. MSL, EM, MSF, and BW analyzed the data. ZZ, XPL, and SAY performed the density functional theory calculations. DK provided the ZrSiS bulk crystals. BW conceived and coordinated the project. MSL, EM, and BW wrote the manuscript with input from all authors.

**ACKNOWLEDGMENT**

This research is supported by the National Research Foundation (NRF) Singapore, under the Competitive Research Programme "Towards On-Chip Topological Quantum Devices" (NRF-CRP21-2018-0001) with further support from the Singapore Ministry of Education (MOE) Academic Research Fund Tier 3 grant (MOE2018-T3-1-002) "Geometrical Quantum Materials", and a Singapore MOE AcRF Tier 2 (Grant No.MOE-T2EP50220-0011). BW acknowledges a Singapore National Research Foundation (NRF) Fellowship (NRF-NRFF2017-11).

**CONFLICT OF INTEREST STATEMENT**

The authors declare no conflict of interest.

**DATA AVAILABILITY**

The data that supports the findings of this study are available from the corresponding author upon reasonable request.